\documentclass[12pt]{iopart}
\begin{document}

\title[Lattice density-functional theory of surface melting]
{Lattice density-functional theory of surface melting: \\
the effect of a square-gradient correction}

\author{
Santi Prestipino\dag
}

\address{\dag\ Istituto Nazionale per la Fisica della Materia
(INFM), UdR Messina, Italy; and
Universit\`a degli Studi di Messina, Dipartimento di Fisica,
Contrada Papardo, 98166 Messina, Italy
}

\begin{abstract}
I use the method of classical density-functional theory in the
weighted-density approximation of Tarazona to investigate the phase
diagram and the interface structure of a two-dimensional lattice-gas
model with three phases -- vapour, liquid, and triangular solid.
While a straightforward mean-field treatment of the interparticle
attraction is unable to give a stable liquid phase, the correct
phase diagram is obtained when including a suitably chosen
square-gradient term in the system grand potential.
Taken this theory for granted, I further examine
the structure of the solid-vapour interface as the triple point is
approached from low temperature.
Surprisingly, a novel phase (rather than the liquid) is found
to grow at the interface, exhibiting an unusually long modulation
along the interface normal. The conventional surface-melting
behaviour is recovered only by artificially restricting the
symmetries being available to the density field.
\end{abstract}


\submitto{\JPCM}

\maketitle


\section{Introduction}

The statistical behaviour of a spatially inhomogeneous classical fluid
(as is the case of the interface between two coexisting bulk phases or
that of a fluid in a confined geometry) is best analysed within the
(classical) density-functional theory (DFT)~\cite{Evans,reviews}.
Nowadays, the DFT method is widely recognized as being one of the
most accurate theoretical tools in statistical mechanics, as also
proved by the increasing number of applications in the scientific
literature~\cite{jpcm}.

Generally speaking, the DFT provides a functional relation between
the thermodynamic grand potential of the system and its local-density
profile $n(x)$. In practice, the exact relation is not known and one
must resort to approximations: typically, good thermodynamic properties
of the solid phase are obtained by using only the structural properties
of the fluid phase as input. For instance, in the so called
``weighted-density approximation'' (WDA)~\cite{Tarazona1,Curtin1},
one maps the free energy $F[n]$ of the inhomogeneous system onto the
free energy of a fluid with a smoothed density $\bar{n}(x)$, which is
related in a non-local way to $n(x)$. Another popular DFT is the
fundamental-measure theory (FMT)~\cite{Rosenfeld} which, by enforcing
dimensional crossover, provides the best theory invented so far for
the hard-sphere solid~\cite{Tarazona2}. Recently, the FMT has been
extended to hard-core lattice gases by Lafuente and Cuesta~\cite{Lafuente}
and to soft-repulsive systems by Schmidt~\cite{Schmidt}.
While both the WDA and the FMT work pretty well for purely repulsive
systems, they must at present be used in tandem with perturbation theory
when the interaction between particles has an attractive component as
well~\cite{Curtin2,Tang,Mederos,Ohnesorge,Sweatman}. In this case,
however, there is no guarantee that the DFT answer will be accurate.

Recently, I have carried out (together with P. V. Giaquinta) a WDA
study~\cite{ldft} of a lattice-gas system with a realistic phase
diagram ({\it i.e.}, one hosting a solid, a liquid, and a vapour
phase)~\cite{Prestipino}, showing the effectiveness of the DFT method
also for lattice problems. However, the theory of Ref.\,\cite{ldft}
uses different prescriptions, in the solid and in the fluid phase,
for the contribution of the interparticle attraction to the free energy,
and this may represent a serious limitation if one is planning to use
the same density functional of the bulk also for the interface problem.
In particular, an interesting situation to analyse would be that of the
solid-vapour interface at a temperature being only slightly below the
triple-point temperature. In this case, the melting of the surface
usually takes place, with the appearance of a thin liquid layer right
at the interface (a successful DFT of this phenomenon for the 3D
continuum case can be found in \cite{Ohnesorge}). As a matter of fact,
we were not able to provide in \cite{ldft} a sharp evidence of the
surface melting since this would rather require that the analytic
form of the (approximate) grand potential of the system be unique
for all phases.

In the present paper, I develop such a unique density functional for
the same lattice model as in Ref.\,\cite{ldft} by the inclusion of a
convenient square-gradient term. In particular, the new theory leads to
a genuine triple point when treating the interparticle attraction as a
mean field, thus being a natural candidate for a practical demonstration
of the surface-melting phenomenon in a lattice system.
The toll of introducing an adjustable parameter into the theory (which,
as a result, partly loses contact with the original model) is fully
compensated by the gain of a density functional which shares with the
exact unknown one a three-minima structure in the triple-point region.
It comes as a surprise, however, that this is not sufficient to produce
an ordinary surface melting. In fact, an unphysical periodic phase
-- an artefact of the theory -- is unexpectedly found to condense at
the solid-vapour interface in place of the liquid. Only by carrying
out a suitably constrained minimization of the density functional,
aimed at washing out the undesired spurious phase, I finally succeed
to obtain the first, at least to my knowledge, DFT description of
complete surface melting in a lattice system.

This paper is organised as follows. After a brief outline, in section 2,
of the lattice DFT, I describe my system and method in section 3, while
presenting results for the bulk and the solid surface in section 4.
Finally, a summary of the main conclusions is given in section 5.

\section{The lattice density-functional theory: a tutorial}

The lattice DFT~\cite{Nieswand,Reinel,ldft} is a general framework for
describing the statistical properties of systems of lattice particles
under the influence of a site-dependent external potential $\epsilon$
or in the presence of a self-sustained inhomogeneity ({\it e.g.} the
density modulation of a periodic solid). The statistical description
is accomplished in terms of the temperature ($T$) and chemical-potential
($\mu$) evolution of the local density $n_x=\left< c_x\right>$, a
grand-canonical average of the occupation number $c_x=0,1$ of site $x$
in the lattice. $n_x$ is a lattice field which, by the
Hohenberg-Kohn-Mermin (HKM) theorem, is in a one-to-one correspondence
with the external field $\mu_x=\mu-\epsilon_x$, allowing one to define
the Legendre transform $F[n]$ of the grand potential $\Omega$ with
respect to the external field. $F[n]$ is the inhomogeneous counterpart
of the Helmholtz free energy (a more detailed account of the DFT method
can be found in \cite{ldft}).

Given an external field $\mu_x$, one defines a sort of generalized
grand potential,
\begin{equation}
\Omega_{\mu}[\rho]=F[\rho]-\sum_x\mu_x\rho_x\,,
\label{eq01}
\end{equation}
depending on two lattice fields, $\rho$ and $\mu$, which are
regarded as being mutually independent. A minimum principle holds for
(\ref{eq01}), saying that $\Omega_{\mu}[\rho]$ attains its minimum for a
density profile $n_x$ which is precisely the one determined by $\mu_x$.
Moreover, this minimum value is nothing but the grand potential $\Omega$
for the given $\mu_x$. In practice, the exact profile of $F[n]$ is not
known (the ideal gas being an outstanding exception), and one should
assign a form to $F[n]$ which is then used for deriving, via
Eq.\,(\ref{eq01}), an approximate grand potential for the system.

To help the choice of $F[n]$, one usually works with the derivatives
of its excess part $F^{\rm exc}[n]=F[n]-F^{\rm id}[n]$, that is with
the one- and two-point direct correlation functions (DCF), which, in
discrete space, read as:
\begin{equation}
c^{(1)}_x[n]=-\beta\frac{\partial F^{\rm exc}[n]}{\partial n_x}
\,\,\,\,\,{\rm and}\,\,\,\,\,
c^{(2)}_{x,\,y}[n]=-\beta\frac{\partial^2 F^{\rm exc}[n]}
{\partial n_x\partial n_y}\,,
\label{eq02}
\end{equation}
where $\beta=1/(k_BT)$. For a fluid system with density $\rho$,
spatial homogeneity imposes $c^{(1)}_x[n]=c_1(\rho)$ and
$c^{(2)}_{x,\,y}[n]=c_2(x-y,\rho)$.
Knowledge of the (two-point) DCF $c_2(x-y,\rho)$ allows one to
obtain both the excess free energy $f^{\rm exc}(\rho)$ of the
fluid and $c_1(\rho)$ (see \cite{ldft}).

Once given the fluid properties, the excess free energy of the
inhomogeneous system can be formally expressed as an integral
of the DCF through:
\begin{eqnarray}
\fl\beta F^{\rm exc}[n]=N\rho\beta f^{\rm exc}(\rho)-c_1(\rho)\sum_x(n_x-\rho)
\nonumber \\
\lo-\sum_{x,\,y}(n_x-\rho)(n_y-\rho)\int_0^1{\rm d}\lambda\,\int_0^{\lambda}
{\rm d}\lambda^{\prime}\,c^{(2)}_{x,\,y}[n_{\lambda^{\prime}}]\,,
\label{eq03}
\end{eqnarray}
where $N$ is the number of lattice sites and
$n_{\lambda x}=\rho+\lambda(n_x-\rho)$. At this point,
approximations are no more eludible since the functional
$c^{(2)}_{x,\,y}[n_{\lambda^{\prime}}]$ is not known. Among the most
popular of these is the WDA~\cite{Tarazona1,Curtin1}, which assumes
\begin{equation}
F^{\rm exc}[n]\approx F^{\rm exc}_{\rm WDA}[n]=
\sum_xn_xf^{\rm exc}(\bar{n}_x)\,,
\label{eq04}
\end{equation}
where the weighted density $\bar{n}_x=\sum_yn_yw(x-y,\bar{n}_x)$.
Equation (\ref{eq04}) is a refinement of the well-known local-density
approximation (LDA), $F^{\rm exc}_{\rm LDA}[n]=\sum_xn_xf^{\rm exc}(n_x)$.
In a homogeneous system, $\bar{n}_x$ is required to be constant and equal
to the local density, implying a normalization for the weight function $w$;
moreover, $w$ must be such that the fluid DCF be also recovered in the
homogeneous limit. This last condition is translated into a well-definite
algorithm, which allows one to express $\bar{n}_x$ in terms of the local
density. For a lattice system, the details of the WDA method can be found
in Ref.\,\cite{ldft} under the further simplifying hypothesis (first
considered by Tarazona~\cite{Tarazona1}) of a weight function being a
second-order polynomial in the density.

The final step in the method is the calculation of the difference
$\Delta\Omega[n]$ between the generalized grand potential of an
inhomogeneous phase ({\it e.g.} a crystalline solid) and the fluid one,
for equal values of $T$ and $\mu$. One finds:
\begin{eqnarray}
\fl\beta\Delta\Omega[n]=\sum_x\left[ n_x\ln\frac{n_x}{\rho}+
(1-n_x)\ln\frac{1-n_x}{1-\rho}\right] +c_1(\rho)\sum_x(n_x-\rho)
\nonumber \\
\lo+\beta F^{\rm exc}[n]-N\rho\beta f^{\rm exc}(\rho)\,.
\label{eq05}
\end{eqnarray}
After minimizing $\Delta\Omega[n]$ over the density field, the
phase-coexistence condition is imposed by the vanishing of this
minimum difference, which implies the same pressure for both phases.

\section{Model and method}

As a case-study for the surface melting, I shall take a {\it two-dimensional}
(2D) (rather than 3D) lattice-gas model with three phases. By this choice, the
forthcoming analysis of the interface problem gets substantially simplified.
Specifically, I consider the t345 model of Ref.\,\cite{Prestipino}
(see Fig.\,1). This is characterized by a hard-core interaction extending
up to second-neighbor sites in the triangular lattice and a pair attraction
ranging from third- to fifth-neighbor sites. The interaction strenghts are
the same as in \cite{ldft}, namely $v_3=-1.5V,v_4=-1.2V$, and $v_5=-V$,
with $V>0$. The maximum
density allowed to this system is $\rho_{\rm max}=0.25$. The same
model but with $v_3=v_4=v_5=0$ is called the t model. The role of
the t model is that of a reference system since, in the following,
the free energy of the inhomogeneous t345 system is going to be
estimated through the perturbation-theory formula~\cite{Evans,ldft}
\begin{equation}
F[n]=F_0[n]+\sum_{x<y}\Delta v(|x-y|)\left< c_xc_y\right> _0\,,
\label{eq06}
\end{equation}
where the subscript 0 refers to the t model and
$\Delta v(|x-y|)=v(|x-y|)-v_0(|x-y|)$ is the departure of the t345
pair potential from the reference one ({\it i.e.}, $\Delta v=0$ inside
the core while $\Delta v=v$ outside). In particular, it is assumed by the
mean-field approximation (MFA) that $\left< c_xc_y\right> _0=n_xn_y$.

In Ref.\,\cite{ldft}, an approximate DCF for the t fluid is obtained
by solving the Ornstein-Zernike relation in the mean-spherical
approximation (MSA).
It turns out that this MSA solution only exists for all fluid densities
$\rho$ up to $\rho=0.21$. Beyond this threshold, the profile of
$f_0^{\rm exc}(\rho)$ is extrapolated as a fourth-order polynomial
in the density.

The free energy of the inhomogeneous t system is expressed by the lattice
counterpart of the WDA. The further inclusion of the mean-field attraction
leads eventually to a density functional for
the t345 model which, however, turns out to be insufficient to produce a
liquid basin in the phase diagram~\cite{ldft}. In particular, the value
of the freezing density at the critical-point temperature happens to be
too small. To overcome this problem, one solution could be to modify
the mean-field functional in such a way as to shift the freezing line
upward in density, while leaving unaltered the liquid-vapour coexistence
locus. To this aim, a simple square-gradient (SG) correction is certainly
effective: it delays the appearance of the solid phase, thus indirectly
promoting the fluid phase. In fact, the SG term has a long tradition,
appearing in many popular phenomenological theories of the interface
structure~\cite{Evans,reviews}.

Indeed, a general plausibility argument for the SG correction can be
provided, at least in case of a density field which is slowly-varying
in space.
As discussed in appendix A, the leading terms in the exact perturbative
expansion of the excess free energy around a uniform density profile
are a LDA free energy plus a SG term.
However, finding out a direct link between this term and the
microscopic-model parameters is a hard problem (see, however, appendix A).
In fact, an even more difficult task could be to present a theoretical
justification, within the perturbation theory, for the appearance of
a SG correction {\it near} the mean-field attraction.
In particular, including a SG correction into the
mean-field functional might lead to overcounting some of the
contributions to the free energy. However, if our concern is mainly
to formulate a phenomenological DFT theory of surface melting, rather
than reproducing a specific model phase diagram, the above objection
is only of minor significance.
Anyway, an {\it ad hoc} effective SG correction can always be
regarded as just renormalizing the mean-field free energy, {\it i.e.},
as a means for healing the crudeness of the mean-field attraction by
taking into account at least part of the higher-order perturbative
corrections. Hence, it makes sense to examine the
following form of free energy:
\begin{eqnarray}
\fl\beta F[n]=\sum_x\left[ n_x\ln n_x+(1-n_x)\ln(1-n_x)\right]
+\sum_xn_x\beta f_0^{\rm exc}(\bar{n}_x)
\nonumber \\
\lo+\frac{1}{2}\sum_{x,\,y}\beta\Delta v(|x-y|)n_xn_y+
\frac{1}{2}\sum_x\beta J(n_x)\sum_{\delta}(n_x-n_{x+\delta})^2\,,
\label{eq07}
\end{eqnarray}
where $\delta$ denotes a nearest-neighbor direction and $J(\rho)>0$
is, at the moment, a still unspecified function of the density and,
possibly, also of the temperature.
In the spirit of an {\it entirely} phenomenological approach, the
virtue of (\ref{eq07}) is to account for the main features of
the t345 phase diagram, see section 4.1; then, the same theory will
be asked to provide also a sound description of the surface-melting
phenomenon.

For the sake of simplicity, a constant value $\gamma V$ is assumed
for $J(\rho)$. With this choice, the DFT free energy of the t345
system in units of $k_BT$ becomes equal, in the infinite-temperature,
$\beta V\rightarrow 0$ limit, to the WDA free energy of the t system
in the same units ({\it i.e.}, the first two terms on the r.h.s. of
Eq.\,(\ref{eq07})). Moreover, note that the functional of the
t model is not influenced by the SG term in Eq.\,(\ref{eq07}). The
value of $\gamma$ follows after arbitrarily fixing the triple-point
density or temperature (see the next section).

In the homogeneous limit, Eq.\,(\ref{eq07}) gives a generalized
grand potential of
\begin{equation}
\fl\frac{\beta\Omega_{\mu}(\rho)}{N}=
\rho\ln\rho+(1-\rho)\ln(1-\rho)+\rho\beta f_0^{\rm exc}(\rho)+
\frac{1}{2}\rho^2\sum_{n=3}^5z_n\beta v_n-\beta\mu\rho\,,
\label{eq08}
\end{equation}
$z_n$ being the coordination number of the $n$-th lattice shell.
The chemical-potential value that makes (\ref{eq08}) stationary
with respect to $\rho$ is:
\begin{equation}
\beta\mu=\ln\frac{\rho}{1-\rho}-c_{1,0}(\rho)+\rho\sum_{n=3}^5z_n\beta v_n\,,
\label{eq09}
\end{equation}
where $c_{1,0}(\rho)$ is the one-point DCF of the t model. It has
been already discussed in Ref.\,\cite{ldft} that the functional
(\ref{eq08}), with $\beta\mu$ as in (\ref{eq09}), shows two different
minima at low temperature, whence two fluid phases exist, liquid and
vapour, whose relative stability changes upon varying the density $\rho$.

The solid phase has a triangular crystal structure that can be
represented by just two numbers, namely the average densities $n_A$
and $n_B$ at the two sublattices $A$ and $B$ of occupied and unoccupied
sites, respectively. The density functional for this phase then reads:
\begin{eqnarray}
\fl\frac{4\beta\Omega_{\mu}(n_A,n_B)}{N}=n_A\ln n_A+(1-n_A)\ln(1-n_A)
+3\left[ n_B\ln n_B+(1-n_B)\ln(1-n_B)\right]
\nonumber \\
\lo+n_A\beta f_0^{\rm exc}(\bar{n}_A)+3n_B\beta f_0^{\rm exc}(\bar{n}_B)
\nonumber \\
\lo+\frac{1}{2}\left[ n_A\sum_{y|x\in A}\beta v(|x-y|)n_y+
3n_B\sum_{y|x\in B}\beta v(|x-y|)n_y\right]
\nonumber \\
\lo+\frac{1}{2}\gamma\beta V\left[ \sum_{\delta|x\in A}(n_A-n_{x+\delta})^2
+3\sum_{\delta|x\in B}(n_B-n_{x+\delta})^2\right] -\beta\mu(n_A+3n_B)\,.
\label{eq10}
\end{eqnarray}

Upon subtracting (\ref{eq08}) from (\ref{eq10}) and substituting (\ref{eq09})
for $\beta\mu$, one finally arrives, after expanding the sums in (\ref{eq10}),
at the following expression for $\Delta\Omega$:
\begin{eqnarray}
\fl\frac{4\beta\Delta\Omega(n_A,n_B)}{N}=n_A\ln\frac{n_A}{\rho}+
(1-n_A)\ln\frac{1-n_A}{1-\rho}+3\left[ n_B\ln\frac{n_B}{\rho}+
(1-n_B)\ln\frac{1-n_B}{1-\rho}\right]
\nonumber \\
\lo+n_A\beta f_0^{\rm exc}(\bar{n}_A)+3n_B\beta f_0^{\rm exc}(\bar{n}_B)
-4\rho\beta f_0^{\rm exc}(\rho)
\nonumber \\
\lo+3\beta v_3n_A^2+(12\beta v_4+6\beta v_5)n_An_B+
(9\beta v_3+12\beta v_4+6\beta v_5)n_B^2
-2\rho^2\sum_{n=3}^5z_n\beta v_n
\nonumber \\
\lo+6\gamma\beta V(n_A-n_B)^2
+\left( c_{1,\,0}(\rho)-\rho\sum_{n=3}^5z_n\beta v_n\right)
(n_A+3n_B-4\rho)\,.
\label{eq11}
\end{eqnarray}

The weighted densities $\bar{n}_A$ and $\bar{n}_B$ are evaluated
from the values of $n_A$ and $n_B$ with the formulae appearing in
appendix A of Ref.\,\cite{ldft}. In practice, the calculation of
$\bar{n}_A$ and $\bar{n}_B$ is carried out in parallel with that
of $n_A$ and $n_B$, {\it i.e.}, by the same iterative procedure
which determines the minimum of $\Delta\Omega$. This could be a
steepest-descent dynamics or, equivalently, the self-consistent
solution of the equations expressing the stationarity condition
for $\Delta\Omega$. These equations are easily found to be:
\begin{eqnarray}
\fl n_A^{-1}=1+\frac{1-\rho}{\rho}
\exp\left\{ c_1(\rho)-\rho\sum_{n=3}^5z_n\beta v_n+
\beta f^{\rm exc}(\bar{n}_A)
+n_A\beta f^{\rm exc\,\prime}(\bar{n}_A)
\frac{\partial\bar{n}_A}{\partial n_A}\right.
\nonumber \\
\lo+\left. 3n_B\beta f^{\rm exc\,\prime}(\bar{n}_B)
\frac{\partial\bar{n}_B}{\partial n_A}
+6\beta v_3n_A+(12\beta v_4+6\beta v_5)n_B+12\gamma\beta V
(n_A-n_B)\right\} \,;
\nonumber \\
\fl n_B^{-1}=1+\frac{1-\rho}{\rho}
\exp\left\{ c_1(\rho)-\rho\sum_{n=3}^5z_n\beta v_n+
\beta f^{\rm exc}(\bar{n}_B)+
\frac{1}{3}n_A\beta f^{\rm exc\,\prime}(\bar{n}_A)
\frac{\partial\bar{n}_A}{\partial n_B}\right.
\nonumber \\
\lo+\left. n_B\beta f^{\rm exc\,\prime}(\bar{n}_B)
\frac{\partial\bar{n}_B}{\partial n_B}
+(4\beta v_4+2\beta v_5)n_A+(6\beta v_3+8\beta v_4+4\beta v_5)n_B+
4\gamma\beta V (n_B-n_A)\right\} \,.
\nonumber \\
\label{eq12}
\end{eqnarray}
I refer the reader again to \cite{ldft} for more information about
the technicalities of the minimization procedure for $\Delta\Omega$.

\section{DFT results}

Taken the functional (\ref{eq11}) for granted, I now review the results
for the bulk properties of the system. Next, I refer on the attempt
of using the same functional also for the solid-vapour interface.

\subsection{The bulk}

First, I summarize the results of the WDA theory for the t model.
The coexisting fluid and solid densities are found to be
$\rho_{\rm f}=0.1335$ and $\rho_{\rm s}=0.1686$, respectively~\cite{ldft}.
The actual values, obtained through the grand-canonical Monte Carlo (MC)
method~\cite{ldft}, are instead $\rho_{\rm f}=0.172(1)$ and
$\rho_{\rm s}=0.188(1)$, indicating that the thermodynamic stability
of the t solid is overestimated by the theory.

Moving to the attractive model, I first seek for the minimum of
(\ref{eq08}), with $\beta\mu$ as in Eq.\,(\ref{eq09}). For low enough
values of the (reduced) temperature $t=k_BT/V$, there are in fact two
distinct points of minimum, the deeper one being associated with the
thermodynamically stable phase. The locus of points $(\rho,t)$ where
the two minima have equal depth is the liquid-vapour coexistence line
(the crosses in Fig.\,2). In particular, the critical point falls at
$\rho_{\rm cr}=0.068(1)$ and $t_{\rm cr}=0.778(1)$.

As far as the solid-fluid equilibrium is concerned, I first
adjust the quantity $\gamma$ so as to obtain a triple point at a
selected temperature $t_{\rm tr}$. It turns out that, in order for
$t_{\rm tr}$ to be {\it e.g.} exactly 0.7, the value of $\gamma$ in
(\ref{eq11}) has to be 0.25191(1). A bit larger value, {\it i.e.},
$\gamma=0.28214(1)$, shifts the triple-point temperature down to 0.6.

The complete DFT phase diagram of the t345 model is plotted in Fig.\,2
for three values of $\gamma$, namely 0 (MFA), 0.25191 ($t_{\rm tr}=0.7$),
and 0.28214 ($t_{\rm tr}=0.6$). The MC simulation data of
Ref.\,\cite{ldft} are shown as asterisks. It clearly appears that
the functional (\ref{eq11}) yields a realistic phase diagram only
in a small window of $\gamma$ values. In Ref.\,\cite{ldft}, the
predicted DFT phase diagram was equally good (Fig.\,2 of \cite{ldft},
open circles), but we were obliged to use distinct DFT functionals
for the solid and the fluid. A minor defect of the present
theory is the unphysically large ($>0.25$) value of the melting
density at low temperature. In fact, though $n_A$ and $n_B$ are
constrained to be smaller than 1, their combination
$n_{\rm s}=(n_A+3n_B)/4$ is not under control during the functional
minimization. In any event, the agreement of the DFT with MC remains
mainly qualitative, also because of the not excellent quality of the
WDA theory for the reference t system.

As already mentioned, the major advantage of (\ref{eq11}) over the
functional considered in Ref.\,\cite{ldft} lies in the fact that
it has a unique expression for all phases of the system.
This is evidenced {\it e.g.} by its profile at the triple-point
temperature: for $\gamma=0.25191$ ({\it i.e.}, $t_{\rm tr}=0.7$),
the three-minima structure of $\Delta\Omega(n_A,n_B)$ is shown
in Fig.\,3, as projected onto the $n_B=0$ plane in the 3D space
spanned by $n_A,n_B$, and $\Delta\Omega$. Upon moving away from
the triple point, the relative depth of the minima changes, thus
making a particular phase of the system more stable than the others.

\subsection{The solid-vapour interface}

A density functional with three minima, like that at Eq.\,(\ref{eq11}),
gives the rather unique opportunity to describe on an equal footing,
{\it i.e.}, within the same theoretical framework, both the bulk and
the surface statistical properties of a many-particle system. Among
these, a prominent place is certainly held by the surface-melting
phenomenon, which opens a sort of imaginary window on the structure of
the underlying generalized grand potential: roughly speaking, it is
the liquid state claiming visibility before thermodynamics gives its
allowance.

Let me consider, for instance, a linear interface running along a
nearest-neighbor direction, called $X$. This interface breaks the
translational symmetry along the perpendicular, vertical direction
$Y$, thus causing the sublattice densities to get a $Y$ dependence.
Very far from the interface, the densities recover the bulk values,
being those of the solid for {\it e.g.} $Y\ll 0$ and those of the
coexisting vapour for $Y\gg 0$. The horizontal layers are labelled
with an index $\lambda$, which increases upon going from solid
to vapour, being zero at the ``centre'' of the interface. To be
specific, those layers where particles are preferentially hosted
in the solid have an odd $\lambda$ value. At variance with the bulk
case, {\it three} sublattices are to be distinguished now, since
different density values are generally expected at the even and at
the odd interstitial sites. I call $C$ the sublattice formed by the
interstitial sites pertaining to the odd layers, and $B$ the other.
Finally, $A$ is the triangular sublattice that is occupied in the
$T=0$ solid.

Then, a rather straightforward adaptation of (\ref{eq10}) to an
interface geometry yields eventually a surplus
$\Sigma[n]=2\beta\Delta\Omega[n]/N_X$ of grand potential
per surface particle, due to the interface, equal to:
\begin{eqnarray}
\fl\Sigma[n]=\sum_{\lambda\,\,{\rm odd}}\left[
n_{A,\lambda}\ln\frac{n_{A,\lambda}}{\rho}+
(1-n_{A,\lambda})\ln\frac{1-n_{A,\lambda}}{1-\rho}+
n_{C,\lambda}\ln\frac{n_{C,\lambda}}{\rho}+
(1-n_{C,\lambda})\ln\frac{1-n_{C,\lambda}}{1-\rho}\right]
\nonumber \\
\lo+2\sum_{\lambda\,\,{\rm even}}\left[
n_{B,\lambda}\ln\frac{n_{B,\lambda}}{\rho}+
(1-n_{B,\lambda})\ln\frac{1-n_{B,\lambda}}{1-\rho}\right]
\nonumber \\
\lo+\left( c_{1,0}(\rho)-\rho\sum_{n=3}^5z_n\beta v_n\right)
\left[ \sum_{\lambda\,\,{\rm odd}}
(n_{A,\lambda}+n_{C,\lambda}-2\rho)+2\sum_{\lambda\,\,{\rm even}}
(n_{B,\lambda}-\rho)\right]
\nonumber \\
\lo+\sum_{\lambda\,\,{\rm odd}}\left[
n_{A,\lambda}\beta f_0^{\rm exc}(\bar{n}_{A,\lambda})+
n_{C,\lambda}\beta f_0^{\rm exc}(\bar{n}_{C,\lambda})
-2\rho\beta f_0^{\rm exc}(\rho)\right]
\nonumber \\
\lo+2\sum_{\lambda\,\,{\rm even}}\left[
n_{B,\lambda}\beta f_0^{\rm exc}(\bar{n}_{B,\lambda})
-\rho\beta f_0^{\rm exc}(\rho)\right]
\nonumber \\
\lo+\frac{1}{2}\sum_{\lambda\,\,{\rm odd}}\left[
n_{A,\lambda}\sum_{y|x\in A,\lambda}n_y\beta\Delta v(|x-y|)+
n_{C,\lambda}\sum_{y|x\in C,\lambda}n_y\beta\Delta v(|x-y|)-
2\rho^2\sum_{n=3}^5z_n\beta v_n\right]
\nonumber \\
\lo+\sum_{\lambda\,\,{\rm even}}\left[ n_{B,\lambda}
\sum_{y|x\in B,\lambda}n_y\beta\Delta v(|x-y|)-
\rho^2\sum_{n=3}^5z_n\beta v_n\right]
\nonumber \\
\lo+\frac{1}{2}\gamma\beta V\left\{ \sum_{\lambda\,\,{\rm odd}}\left[
\sum_{\delta|x\in A,\lambda}(n_{A,\lambda}-n_{x+\delta})^2+
\sum_{\delta|x\in C,\lambda}(n_{C,\lambda}-n_{x+\delta})^2\right] \right.
\nonumber \\
\lo+\left. 2\sum_{\lambda\,\,{\rm even}}
\sum_{\delta|x\in B,\lambda}(n_{B,\lambda}-n_{x+\delta})^2\right\} \,.
\label{eq13}
\end{eqnarray}
To save space, some of the sums in the above Eq.\,(\ref{eq13}) are left
as indicated; however, expanding them is not particularly difficult,
it is just a matter of carefully looking at the lattice geometry.

I have considered the case $\gamma=0.25191$ and a number of temperature
values in the range from 0.6 to $t_{\rm tr}=0.7$. A slab of interface
generally consisted of 160 layers, from $\lambda=-80$ to $\lambda=79$,
with fixed values of the sublattice densities outside this range. The
extension of the slab in the $X$ direction is virtually infinite. The
interface is prepared by always assuming an inverted-exponential,
$(1+\exp[(\lambda+1/2)/l])^{-1}$ modulation~\cite{ldft} which is
then relaxed until a minimum of $\Sigma[n]$ is found.
Disappointingly, however, the final looking of the interface
was never as expected. A typical outcome is depicted in Fig.\,4, which
refers to $t=0.65$. We can see that the mean density of a layer,
$n_{\lambda}$, shows a curious ten-layer-long periodic motif for
$\lambda>0$, which interrupts at the right border of the picture for
the existence of a fixed boundary. Even more serious is the fact that
the minimum of $\Sigma[n]$ turns out to be {\it negative}, signalling
that there is something wrong with this interface.

To clarify such things better, I have cut out a piece of this periodic
motif and made it infinite in both $Y$ directions. The resulting phase
does, in fact, overcome in stability both the solid and the vapour.
Obviously, such a weird evidence is totally unphysical, being a
spurious result of the DFT, as founded on Eq.\,(\ref{eq13}).
It is worth noticing that, in this periodic phase, the local values
of the system density are different at the three sublattices $A$, $B$,
and $C$, {\it i.e.}, this phase can only exist in an interface.
Nonetheless, the very occurrence of this oddity inevitably casts
a shadow on the real quality of the functional (\ref{eq07}), which
should then be re-examined in spite of its capability of accounting
for the bulk properties of the t345 model.

Two different routes are now opening to our consideration: the first
is simply to reject the functional (\ref{eq07}) as unphysical. Another
solution is however viable, which is to insist on (\ref{eq07}), trying
to see whether it is possible to unveil the normal surface-melting
behaviour by properly obstructing the manifestation of the undesired
phase. Obviously, this is not completely orthodox since, in a sense,
a sort of external field is being introduced into the problem.

If we decide to pursue this second route anyway, one can try the following.
Rather than seeking for the absolute minimum of $\Sigma[n]$, the search
is restricted to just those fields $n_x$ that, at a particular $C$ site,
take the same value as in one of the two closest $B$ sites. Precisely,
I require that $n_{C,\lambda}=n_{B,\lambda-1}$, a choice which is
consistent with the boundary values but incompatible with the unphysical
periodic phase. Whether this is sufficient to promote the appearance of
the liquid at the middle of the interface is a matter of debate that
can only be fixed numerically.

In order to carry out the constrained minimization of $\Sigma[n]$
along the guidelines presented above, Eq.\,(\ref{eq13}) must be
modified by simply substituting $n_{B,\lambda-1}$ for $n_{C,\lambda}$
everywhere it occurs. A more subtle, but absolutely crucial, technical
point is relative to the weighted densities: it follows from their
very definition that there is actually no simple relation between
$\bar{n}_{C,\lambda}$ and $\bar{n}_{B,\lambda-1}$, even when
$n_{C,\lambda}=n_{B,\lambda-1}$. This is due to the fact that,
owing to the $Y$ dependence of the sublattice densities, the
neighborhood of a $C$ site does {\it not} look the same as that
of a $B$ site, and this causes $\bar{n}_{C,\lambda}$ and
$\bar{n}_{B,\lambda-1}$ to be, at least in principle, different.
In the end, the amended $\Sigma[n]$ functional reads as:

\begin{eqnarray}
\fl\Sigma[n]=\sum_{\lambda\,\,{\rm odd}}\left[
n_{A,\lambda}\ln\frac{n_{A,\lambda}}{\rho}+
(1-n_{A,\lambda})\ln\frac{1-n_{A,\lambda}}{1-\rho}\right]
+3\sum_{\lambda\,\,{\rm even}}\left[
n_{B,\lambda}\ln\frac{n_{B,\lambda}}{\rho}+
(1-n_{B,\lambda})\ln\frac{1-n_{B,\lambda}}{1-\rho}\right]
\nonumber \\
\lo+\left( c_{1,0}(\rho)-\rho\sum_{n=3}^5z_n\beta v_n\right)
\left[ \sum_{\lambda\,\,{\rm odd}}
(n_{A,\lambda}-\rho)+3\sum_{\lambda\,\,{\rm even}}
(n_{B,\lambda}-\rho)\right]
\nonumber \\
\lo+\sum_{\lambda\,\,{\rm odd}}\left[
n_{A,\lambda}\beta f_0^{\rm exc}(\bar{n}_{A,\lambda})+
n_{B,\lambda-1}\beta f_0^{\rm exc}(\bar{n}_{C,\lambda})
-2\rho\beta f_0^{\rm exc}(\rho)\right]
\nonumber \\
\lo+2\sum_{\lambda\,\,{\rm even}}\left[
n_{B,\lambda}\beta f_0^{\rm exc}(\bar{n}_{B,\lambda})
-\rho\beta f_0^{\rm exc}(\rho)\right]
\nonumber \\
\lo+\frac{1}{2}\sum_{\lambda\,\,{\rm odd}}\left\{ n_{A,\lambda}\left[
2\beta v_3(n_{A,\lambda-2}+n_{A,\lambda}+n_{A,\lambda+2})+
2\beta v_4(2n_{B,\lambda-3}+n_{B,\lambda-1}+2n_{B,\lambda+1}+
n_{B,\lambda+3})\right. \right.
\nonumber \\
\lo+\left. \left. 2\beta v_5(n_{B,\lambda-3}+n_{B,\lambda-1}+
n_{B,\lambda+3})\right] -\rho^2\sum_{n=3}^5z_n\beta v_n\right\}
\nonumber \\
\lo+\sum_{\lambda\,\,{\rm even}}\left\{ n_{B,\lambda}\left[
3\beta v_3(n_{B,\lambda-2}+n_{B,\lambda}+n_{B,\lambda+2})\right. \right.
\nonumber \\
\lo+\beta v_4(n_{B,\lambda-4}+n_{A,\lambda-3}+4n_{B,\lambda-2}+
2n_{A,\lambda-1}+2n_{B,\lambda}+n_{A,\lambda+1}+4n_{B,\lambda+2}+
2n_{A,\lambda+3}+n_{B,\lambda+4})
\nonumber \\
\lo+\left. \left. \beta v_5(n_{B,\lambda-4}+n_{A,\lambda-3}+n_{B,\lambda-2}+
2n_{B,\lambda}+n_{A,\lambda+1}+n_{B,\lambda+2}+n_{A,\lambda+3}+
n_{B,\lambda+4})\right] \right.
\nonumber \\
\lo-\left. \frac{3}{2}\rho^2\sum_{n=3}^5z_n\beta v_n\right\}
+\gamma\beta V\left\{ \sum_{\lambda\,\,{\rm odd}}\left[
3(n_{A,\lambda}-n_{B,\lambda-1})^2+(n_{A,\lambda}-n_{B,\lambda+1})^2
\right] \right.
\nonumber \\
\lo+\left. \sum_{\lambda\,\,{\rm even}}\left[
(n_{B,\lambda}-n_{A,\lambda-1})^2+2(n_{B,\lambda}-n_{B,\lambda-2})^2+
(n_{B,\lambda}-n_{A,\lambda+1})^2\right] \right\} \,.
\label{eq14}
\end{eqnarray}

I omit to indicate the explicit expression of the weighted densities
and of their derivatives, since they would need too much space to be
specified here.

Using the functional (\ref{eq14}), I have carried out a new series of
minimizations, for various temperature values. With great pleasure,
the thermal evolution of the interface structure is now as expected
in a 3D system with a complete surface melting, see Fig.\,5, though
dissimilar to the behaviour of the actual (2D) t345 model where the
layering of the liquid is much more pronounced, cf. Fig.9 (top) of
Ref.\,\cite{ldft}. Moreover, the present evidence is also superior
to that found in \cite{ldft} by using different functionals for
the fluid and the solid. There, the thickness of the molten layer
was too small even very close to the triple-point temperature.

The slab used in the present optimizations was a
hundred layers wide, from $\lambda=-50$ to $\lambda=49$, but only a
slice of this is shown in Fig.\,5. In Fig.\,6, I report the detailed
profile across the interface of the sublattice densities for two distinct
values of $t$ close to $t_{\rm tr}$, namely $t=0.69$ and $t=0.699$.
By looking at Fig.\,5, we see that the width of the liquid layer grows
with regularity as the triple-point
temperature is approached from below, corresponding to the liquid phase
becoming more and more stable. To estimate this width as a function of
temperature, I use the following three-parameter, double-exponential fit:
\begin{eqnarray}
\fl n_{A,\lambda}=\rho+\frac{n_A-\rho_{\rm l}}{1+\exp\left[
(\lambda+1/2+L)/l_1\right] }+
\frac{\rho_{\rm l}-\rho}{1+\exp\left[ (\lambda+1/2-L)/l_2\right] }\,,
\,\,\,{\rm for}\,\,{\rm odd}\,\,\lambda\,\,{\rm only}\,;
\nonumber \\
\fl n_{B,\lambda}=\rho+\frac{n_B-\rho_{\rm l}}{1+\exp\left[
(\lambda+1/2+L)/l_1\right] }+
\frac{\rho_{\rm l}-\rho}{1+\exp\left[ (\lambda+1/2-L)/l_2\right] }\,,
\,\,\,{\rm for}\,\,{\rm even}\,\,\lambda\,\,{\rm only}\,;
\nonumber \\
\fl n_{C,\lambda}=n_{B,\lambda-1}\,,
\,\,\,{\rm for}\,\,{\rm odd}\,\,\lambda\,\,{\rm only}\,.
\label{eq15}
\end{eqnarray}
In Eq.\,(\ref{eq15}), $\rho_{\rm l}$ is the density of the (metastable)
liquid coexisting with the vapour at the given temperature $t<t_{\rm tr}$,
$2L$ is the nominal width of the liquified part of the interface, while
$l_1$ and $l_2$ are a measure of the width of the two transition regions
being centred at $-L-1/2$ and $L-1/2$, respectively. The values of the
fitting parameters are obtained by substituting the {\it ansatz}
(\ref{eq15}) into Eq.\,(\ref{eq14}) and requiring it to be minimum.
It turns out that this minimum value is never appreciably far from
the absolute minimum of $\Sigma[n]$ as being determined independently.

For $t=0.65,0.69$, and $0.699$, I find $L=2.67,4.61$, and $8.60$,
respectively (cf. Fig.\,5). Hence, while at low temperature it is
more convenient for the solid-liquid and liquid-vapour interfaces to
stay bound together, slightly below the triple point the free-energy
cost of two well-separated interfaces is smaller. In some
phenomenological theories of the surface melting, this effect
is mimicked through the introduction of an effective repulsion
between the two interfaces. Furthermore, while $l_1\simeq 1.0$ shows
only a small dependence on temperature, the increase of $l_2$ with $t$
is more conspicuous, being $1.42,1.81$, and $2.86$ for the above cited
$t$ values. At these same temperatures, the minimum $\Sigma[n]$
turned out to be $1.10162,0.74472$, and $0.65504$, respectively.
In particular, the latter value would practically account for the
overall cost of two separate solid-liquid and liquid-vapour
interfaces at $t=t_{\rm tr}$. I do not try to deduce from the
above numbers an empirical relation between $2L$ and $t$, {\it e.g.}
with the aim at resolving a logarithmic from a power-law behaviour.
In fact, owing to a necessarily imperfect estimate of the
coexistence conditions within the DFT, it was actually impossible to
approach the triple point more closely than 0.001 in $t$. This
notwithstanding, I do not find any reason to doubt that $L$ would
grow to infinity for $t\rightarrow t_{\rm tr}^-$.

Summing up, the surface melting is observed, within the functional
(\ref{eq11}), only at the price of artificially restricting the
symmetries that are available to the density field. Otherwise,
in fact, a spurious periodic phase will wet the solid, which is
actually more stable than the solid and vapour themselves.

\newpage
\section{Conclusions}

The usual realm of applications of the classical DFT comprises hard-core
systems and soft-repulsive ones. Besides those cases, the only existing
general DFT method is, at present, the density-functional perturbation
theory which, however, can be quantitatively inaccurate or, even, predict
erroneous phase diagrams.

In this paper, I have used the lattice analogue of the perturbative DFT
in order to analyse the phase behaviour and the interface structure of
a 2D lattice-gas model with a pair interaction consisting of a hard
core plus an attractive tail. From previous studies, this model is
known to exhibit a solid, a liquid, and a vapour phase.

In order to obtain a phase diagram with three phases, I have
considered a phenomenological density functional which, besides
treating the interparticle attraction as a mean field,
also contains a tunable square-gradient correction. In fact, the
very unique request to this theory was to yield a generalized grand
potential for an inhomogeneous phase that could exactly turn, in the
homogeneous limit, into the fluid functional. This requirement is
crucial for getting a consistent description of the interface
between the solid and a fluid phase.

After producing a good theory for the bulk, I have adapted the
same density functional to the geometry of a solid-vapour interface.
The aim was to observe, under equilibrium conditions, the melting of
the solid surface, {\it i.e.}, the growth of a liquid layer between
the solid and the vapour as the triple point is approached from low
temperature.

Surprisingly, the minimization of the interface functional unveils
instead a periodic phase which is totally unphysical, being actually
an artefact of the theory. In order to recover a more conventional
behaviour, I have carried out a constrained functional minimization
which, while not affecting the liquid, makes it unlikely for more exotic
phases to appear. By this stratagem, and nothwistanding the partial
arbitrariness that is implicit in such a conditioned minimization,
in the end I arrive at a fairly realistic description of the complete
surface melting of a lattice system -- a result which, in my opinion,
is an interesting illustrative example of the risks which may accompany
the use of a phenomenological DFT functional. Anyway, I do not think
that the {\it ad hoc} expedient that was considered here can be of
help for investigating, in the absence of a good-quality functional,
more delicate features like {\it e.g.} layering phase transitions and
roughening.

\newpage
\appendix
\section{Estimate of the square-gradient correction}

In this appendix, I show how to adapt to a discrete space an argument
originally due to Evans about the estimate of the ``optimal'' SG
correction to a LDA functional~\cite{Evans}.

Let the following excess free energy be considered,
\begin{equation}
F^{\rm exc}[n]=\sum_xn_xf^{\rm exc}(n_x)
+\frac{1}{2}\sum_xJ(n_x)\sum_{\delta}(n_x-n_{x+\delta})^2\,,
\label{a01}
\end{equation}
which is a LDA free energy supplemented with a SG term.
This form of $F^{\rm exc}[n]$ is not precisely the one which I work
with in section 3 for demonstrating the surface-melting phenomenon.
Actually, my intention here is not to provide a full justification
of Eq.\,(\ref{eq07}); rather, I just want to give a flavour of the
physical status of an SG correction.

To state the problem precisely, the function $J(\rho)$ in
Eq.\,(\ref{eq07}) will be calculated from the requirement that
at least the second-order expansion of (\ref{a01}) around a fluid
of density $\rho$ may agree with the exact one,
\begin{equation}
\fl F^{\rm exc}[n]=N\rho f^{\rm exc}(\rho)-\frac{1}{\beta}c_1(\rho)
\sum_x\Delta n_x -\frac{1}{2\beta}
\sum_{x,\,y}c_2(x-y,\rho)\Delta n_x\Delta n_y+\ldots
\label{a02}
\end{equation}
The field $\Delta n_x=n_x-\rho$ is a measure of the (admittedly
small) deviation of the system from homogeneity.

Since
\begin{equation}
c_1(\rho)=-\beta f^{\rm exc}(\rho)-\rho\beta f^{{\rm exc}\,\prime}(\rho)\,,
\label{a03}
\end{equation}
the expansion of the LDA term in (\ref{a01}) is as follows:
\begin{equation}
\fl\sum_xn_xf^{\rm exc}(n_x)=N\rho f^{\rm exc}(\rho)-\frac{1}{\beta}c_1(\rho)
\sum_x\Delta n_x-\frac{1}{2\beta}c_1^{\prime}(\rho)\sum_x\Delta n_x^2+\ldots
\label{a04}
\end{equation}
To proceed further, it is convenient to work in Fourier space.
I take the Fourier transform of a lattice field $f_x$ to be
$\tilde{f}_q=\sum_xf_x\exp(-{\rm i}q\cdot x)$ (see the precise
definition of the $q$ vectors in Ref.\,17 of \cite{ldft}). The
convolution theorem first yields:
\begin{equation}
\sum_x\Delta n_x^2=\frac{1}{N}\sum_q\tilde{\Delta n}_q\tilde{\Delta n}_{-q}\,;
\label{a05}
\end{equation}
similarly, the fully quadratic, leading SG correction becomes as follows:
\begin{equation}
\fl\frac{1}{2}J(\rho)\sum_{x,\,\delta}(n_x-n_{x+\delta})^2\equiv
J(\rho)\sum_{x,\,y}W(x-y)\Delta n_x\Delta n_y=
J(\rho)\frac{1}{N}\sum_q\tilde{W}_q\tilde{\Delta n}_q\tilde{\Delta n}_{-q}\,.
\label{a06}
\end{equation}
The function $W$ in Eq.\,(\ref{a06}) is lattice-dependent; for {\it e.g.}
a $d$-dimensional hypercubic lattice with unit lattice constant,
$W(x-y)=2d\delta_{x,\,y}-\delta_{|x-y|,1}$ and
$\tilde{W}_q=2d[1-(1/d)\sum_{\alpha=1}^d\cos q_{\alpha}]$. Finally,
\begin{equation}
\sum_{x,\,y}c_2(x-y,\rho)\Delta n_x\Delta n_y=
\frac{1}{N}\sum_q\tilde{c_2}(q,\rho)\tilde{\Delta n}_q\tilde{\Delta n}_{-q}\,.
\label{a07}
\end{equation}

Upon comparing Eqs.\,(\ref{a04}), (\ref{a05}), and (\ref{a06}) with
Eq.\,(\ref{a07}), one would be led to identifying
$c_1^{\prime}(\rho)-2\beta J(\rho)\tilde{W}_q$ with $\tilde{c_2}(q,\rho)$.
However, it is clear that there is no chance for these two quantities to
be exactly equal, since $J(\rho)$ does not depend on $q$. To fix this
problem, one should make some further assumption on the $\Delta n_x$.
For instance, one can require that, besides being small, $\Delta n_x$
also varies slowly in space. In this case, only the long-wavelength Fourier
components of $\Delta n_x$ are important, implying that what really
matters in fact is the matching of the above two quantities in the
small-$q$ limit. In particular, the LDA free energy ({\it i.e.},
$J(\rho)=0$) does exactly reproduce only the $q=0$ Fourier component
in Eq.\,(\ref{a07}), since $c_1^{\prime}(\rho)=\tilde{c_2}(0,\rho)$.

At second order in $q$,
\begin{equation}
\tilde{c_2}(q,\rho)=\sum_xc_2(x,\rho)\left( 1-{\rm i}q\cdot x-\frac{1}{2}
(q\cdot x)^2+\ldots\right) \,.
\label{a08}
\end{equation}
The first term $\sum_xc_2(x,\rho)=\tilde{c_2}(0,\rho)$.
The second term is zero for a lattice with inversion symmetry ({\it i.e.},
one with $c_2(x,\rho)=c_2(-x,\rho)$). The third term is:
\begin{equation}
\fl-\frac{1}{2}\sum_xc_2(x,\rho)(q\cdot x)^2=
-\frac{1}{2}q^2\sum_xx^2c_2(x,\rho)(\hat{q}\cdot\hat{x})^2=
-\frac{1}{2d}q^2\sum_xx^2c_2(x,\rho)\equiv d_2(\rho)q^2\,,
\label{a09}
\end{equation}
having denoted unit vectors by a hat. Observe that, in the last step,
the axial symmetry of the lattice with respect to each coordinate axis
has been assumed. Hence, the expansion of $\tilde{c_2}(q,\rho)$ in powers
of $q$ begins as $\tilde{c_2}(0,\rho)+d_2(\rho)q^2$.

Since $\tilde{W}_q=q^2+o(q^2)$ for a hypercubic lattice, the expression
of $J(\rho)$ for this lattice is:
\begin{equation}
J(\rho)=-\frac{1}{2\beta}d_2(\rho)=\frac{1}{4\beta d}\sum_xx^2c_2(x,\rho)\,.
\label{a10}
\end{equation}
For a triangular lattice, $\tilde{W}_q=(3/2)q^2+o(q^2)$ while $d_2(\rho)$
is like in Eq.\,(\ref{a09}) with $d=2$. Hence, $J(\rho)$ is still given
by Eq.\,(\ref{a10}), but with $d=3$ (the rationale being that, for the
triangular lattice, the nearest-neighbor sites are six, like for the
cubic lattice).

To conclude, I make some comments on the result (\ref{a10}). For the
t model, the non-zero, core values of $c_2(x,\rho)$ are all negative,
hence $J(\rho)<0$. This means that the SG correction is meaningless
for this model. Different is the case of the t345 model, where the
positive tail of the DCF would eventually make $J(\rho)>0$.
However, the exact $c_2(x,\rho)$ is not known, and this actually
makes Eq.\,(\ref{a10}) rather useless.

\newpage
\begin{center}
{\bf Figure Captions} \vspace{5mm}
\end{center}
\begin{description}
\item[{\bf Fig.\,1 :}] The triangular-lattice model under consideration
(schematic): excluded sites ($\times$) and attractive sites (numbered dots)
are separately shown for a particle sitted at the centre of the picture
(large black dot).
The attractive interaction reaches the fifth neighbours, no interaction
is felt beyond this distance (small dots). The Hamiltonian is of the form
$H=\sum_{i<j}v(|i-j|)c_ic_j$, where $v(|i-j|)=+\infty$ when simultaneous
occupation of sites $i$ and $j$ is forbidden. In the text, $v_n$ denotes
the value of $v(|i-j|)$ for a pair of $n$-th neighbours, whereas $z_n$ is
the number of $n$-th neighbours ($z_1=z_2=z_3=6$, $z_4=12$, and $z_5=6$).
In this work, the interaction strengths were $v_3=-1.5V$, $v_4=-1.2V$,
and $v_5=-V$ (with $V>0$). For these $v_n$, the phase diagram of the
lattice model is the canonical one, with a solid, a liquid, and a
vapour phase (see Ref.\,\cite{Prestipino}).

\item[{\bf Fig.\,2 :}] Phase diagram of the t345 model as drawn from
the density functional (\ref{eq11}), using the t model (MSA+WDA) as a
reference and representing the interparticle attraction as a mean-field
perturbation. Various solid-fluid coexistence lines are shown: $\gamma=0$
($\triangle$ and dotted lines)~\cite{ldft}, $\gamma=0.25191$ ($\bigcirc$
and continuous lines), and $\gamma=0.28214$ ($\opensquare$ and continuous
lines). The liquid-vapour coexistence locus is marked with crosses.
For comparison, I have reported as asterisks joined by dashed
lines some MC data points for the $48\times 48$ lattice (the errors
affecting these points are of the same size as the symbols). The arrows
pointing downwards mark the WDA densities of the coexisting fluid and
solid in the t model. The other arrows point to the MC estimates for
the same quantities.

\item[{\bf Fig.\,3 :}] The function $\Delta\Omega(n_A,n_B)$ at
Eq.\,(\ref{eq11}) is plotted vs. $n_A$, for various $n_B$ values
ranging from 0 to 1 (both sublattice densities are incremented in
steps of $5\times 10^{-3}$). Here, $\gamma=0.25191$ and $t=t_{\rm tr}=0.7$.
Straight lines are drawn through the points as a guide to the eye.
The vapour and liquid densities are $\rho_{\rm v}=0.0248$ and
$\rho_{\rm l}=0.1282$, respectively. The coordinates of the ``solid''
minimum are $n_A=0.94767$ and $n_B=0.02765$, corresponding to a solid
density of $\rho_{\rm s}=0.2577$.

\item[{\bf Fig.\,4 :}] DFT profile of the mean density $n_{\lambda}$ at
the $\lambda$-th layer, defined as $(n_{A,\lambda}+n_{C,\lambda})/2$
for odd $\lambda$; as $n_{B,\lambda}$ for even $\lambda$.
What is actually plotted is the outcome of the minimization of
(\ref{eq13}) for $\gamma=0.25191$ and $t=0.65$. During the
minimization process, it is noted that the vapour disappears
progressively in favour of a novel phase with a period of ten
layers. The automatic search of the minimum $\Delta\Omega$ comes
to a stop when this periodic phase reaches the border of the slab.

\item[{\bf Fig.\,5 :}] DFT results from the minimization of
functional (\ref{eq14}) for $\gamma=0.25191$ ($t_{\rm tr}=0.7$):
the optimum mean-density profile across the solid-vapour interface
is plotted for three values of $t$, namely $0.65,0.69$, and $0.699$.
To help the eye, in each panel a continuous line has been drawn
through the points. As is quite clear from the pictures, the
typical surface-melting behaviour of a 3D simple fluid is
derived from Eq.\,(\ref{eq14}).

\item[{\bf Fig.\,6 :}] Layer-by-layer evolution of $n_{A,\lambda}$
(for odd $\lambda$ only), $n_{B,\lambda}$ (for even $\lambda$ only),
and $n_{C,\lambda}=n_{B,\lambda-1}$ (for odd $\lambda$ only) for the
two same interface profiles of highest temperature as in Fig.\,5. Near
$\lambda=0$, the plateau of $n_{A,\lambda}$ and the maximum of both
$n_{B,\lambda}$ and $n_{C,\lambda}$ are clear signs of a local
liquid-like behaviour.
\end{description}
\end{document}